\documentstyle[aps,prl,floats,twocolumn,psfig]{revtex}

\def\JPhysC#1#2#3{J.\ Phys.\ C {\bf {#1}}, {#2} (19{#3})}

\def\JETPL#1#2#3{JETP Lett.\ {\bf {#1}}, {#2} (19{#3})}

\def\PRB#1#2#3{Phys.\ Rev.\ B {\bf {#1}}, {#2} (19{#3})}
\def\PRL#1#2#3{Phys.\ Rev.\ Lett.\ {\bf {#1}}, {#2} (19{#3})}

\def\ZhETF#1#2#3#4#5{Zh.E.T.F.\ {\bf {#1}}, {#2} (19{#3}) [JETP {\bf {#4}}, 
{#5} (19{#3})]}

\def\Science#1#2#3{Science {\bf {#1}}, {#2} (19{#3})}

\def\ModPL#1#2#3{Mod.\ Phys.\ Lett.\ {\bf {#1}}, {#2} (19{#3})}

\def\MicroE_Eng#1#2#3{Microelect.\ Eng.\ {\bf {#1}}, {#2} (19{#3})}

\def\RMP#1#2#3{Rev.\ Mod.\ Phys.\ {\bf {#1}}, {#2} (19{#3})}

\def\gap{\scriptsize${\stackrel{\textstyle _>}{_\sim}} $\normalsize \  }
\def\lap{\scriptsize${\stackrel{\textstyle _<}{_\sim}} $\normalsize \  }

\def\UNC{Dept.\ of Physics and Astronomy, The University of North 
Carolina at Chapel Hill, Chapel Hill, NC 27599-3255}
\def\IBM{IBM Research Division, T.J.\ Watson Research Center, Yorktown
Heights, NY 10598}
\def\FSU{National High Magnetic Field Laboratory, 
Florida State University, Tallahassee, FL 32306}

\begin{document}

\twocolumn[\hsize\textwidth\columnwidth\hsize\csname @twocolumnfalse\endcsname

\title{Absence of Localization in Certain Field Effect Transistors}

\author{S.\ Washburn$^1$, Dragana Popovi\'{c}$^2$, K.P.\ Li $^1$ 
and A.B.\ Fowler$^{3}$}
\address{$^1$\UNC \\ 
$^2$\FSU \\
$^3$ \IBM }

\date{\today}
\maketitle

\begin{abstract}

We review some experimental and theoretical results on the
metal-to-insulator transition (MIT) observed at zero magnetic field
($B=0$) in several two-dimensional electron systems (2DES).
Scaling of the conductance and magnetic field dependence of the
conductance provide convincing evidence that the MIT is driven by
Coulomb interactions among the carriers and is dramatically sensitive
to spin polarization of the carriers.  
\end{abstract}

%
%
%
]

\section{Introduction}

Landauer's early work on scattering of particles and waves in random
media\cite{RL} is still important today, because after decades of
work, the physics of transport in disordered systems is still keeping
us entertained with unexpected results.  The generic ``conductivity''
that arises from ham-handedly averaging microscopic details and
sweeping Coulomb interactions under the rug of effective mass is a
poor description of transport in many circumstances.  In the past two
decades there have been several discoveries in experiments (much of it
spurred or explained theoretically by Landauer and his co-workers)
that have flown dramatically in the face of the classical conductivity.
Response from quantum mechanically phase-coherent carrier excitations
have appeared in studies of metals and semiconductors in both the
relatively clean and very dirty limits of impurity scattering.  In
high magnetic fields the re-writing of Ohm's law has been especially
dramatic.  Noise from fluctuations in the transport parameters have
behaved in ways that contradict the simple models that lead to the
generic resistivity.\cite{RL_noise}  
Finally, Coulomb interactions are appearing as
large perturbations of the non-interacting carrier models that have
dominated thinking for two decades.

\begin{figure}
\centerline{\psfig{figure=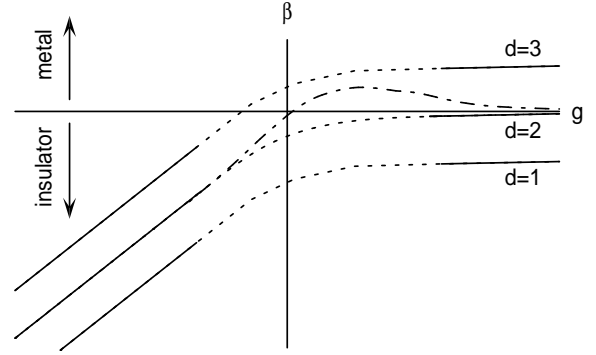,width=3in}}
\vspace{0.2cm}
\caption{
Schematic theoretical plot of $\beta$ with calculated results (solid lines) 
for noninteracting particles in the limits $\beta \gg 1$ or $\beta \ll 1$ and
plausibility arguments (dotted lines) connecting them.  Alternative versions
(dot-dash line) for $d=2$ have been proposed for interacting particles.
\label{betaplot}
}
\end{figure}

Since the appearance of the scaling theory for non-interacting
(Fermi-liquid) particles\cite{gang}, most have accepted the
result that there is no minimum metallic conductivity in three
dimensions (3D) and no metallic behavior at all in two dimensions
(2D).  The non-interacting models were developed in terms of a scaling
function
\begin{equation}
\beta ~ = ~ \frac{d(\ln g)}{d(\ln L)}
\label{beta_defn}
\end{equation}
that related dimensionless conductance $g=G/(e^2/h)$ ($G$ is the
conductance of the sample in $1/\Omega$) to the size $L$ of the
(hyper-cube) sample.  For very low conductance ($g \ll 1$), the
transport is exponentially reduced as $L$ increases because the
carriers are localized and must hop or tunnel from site to site.  For
$g \gg 1$, Ohm's law is assumed to hold: $g \propto L^{d-2}$ for cubes
of dimension $d$.  In weak disorder, perturbative calculations from
non-interacting models of carrier transport lead to reduction of
$\beta$ indicating a (perfectly plausible) trend toward lower
conductance and eventually to an insulator as the amount of disorder
increased (illustrated in figure \ref{betaplot}).
According to this picture, 
a 3D system is a conductor if the Fermi energy $E_F$ is larger
than the characteristic amplitude of the disorder potential $W$.  This
criterion is equivalent to the Ioffe-Regel criterion that $k_F l > 1$,
where $k_F$ is the Fermi wave-number and $l$ is the mean free path.
In 2D, there is no metallic behavior for any value of $W \ne
0$.  That is,  the proposed 2D scaling function $\beta < 0$.
Both proclamations are reasonable at first glance, 
but in both cases the
Coulomb interactions among the carriers have largely been swept 
beneath the rug of effective mass and (often ignored) Fermi liquid
parameters.  
Subsequent work, however, has brought to question the validity
of these simple ideas. In particular, 
the absence of a MIT in 2D has been disputed by
Finkelstein\cite{Finkelstein}, who has emphasized that corrections
from interacting particles become important as $T=0$ is approached.
These and more recent \cite{newgang} considerations of interacting systems,
have suggested that $\beta$ can change sign for $d=2$, {\it i.e.} that
there can exist a 2D MIT.

In fact, a wide variety of two-dimensional systems exhibit transitions
like the MIT.  Kosterlitz-Thouless-Berenshinskii transitions\cite{KTB}
in 2D superfluids is an old example.  There are more recent examples
of proper conductor-insulator transitions in two-dimensional
electrical transport problems.  Granular films of superconducting
materials exhibit a transition from insulator to
superconductor as the film thickness increases\cite{SIT_Exp}. The
transition is driven by the superconducting fluctuations in
competition with the Coulomb (charging) interactions between the
grains of material\cite{SIT_Th}. In oxide superconductors, the 2D
sheets of oxygen atoms lead to a highly directional order parameter
and, depending on doping level, exhibit a superconductor to insulator
transition somewhat like that in granular films.\cite{HiTc} Between
plateaux in the quantized Hall effect, which demands very high
magnetic field $B$ such that the Landau energy $\hbar \omega_c \gg
E_F$, there is another 2D MIT manifest as a reversal of the
temperature coefficient of the resistivity
$\rho$.\cite{Shahar,Diorio}.  All of these examples appear in rather
exotic circumstances (Landau quantization or Cooper pairing), but
there is a manifestation of a 2D MIT in garden-variety conductors and
that transition is the subject of the present work.

\begin{figure}
\centerline{\psfig{figure=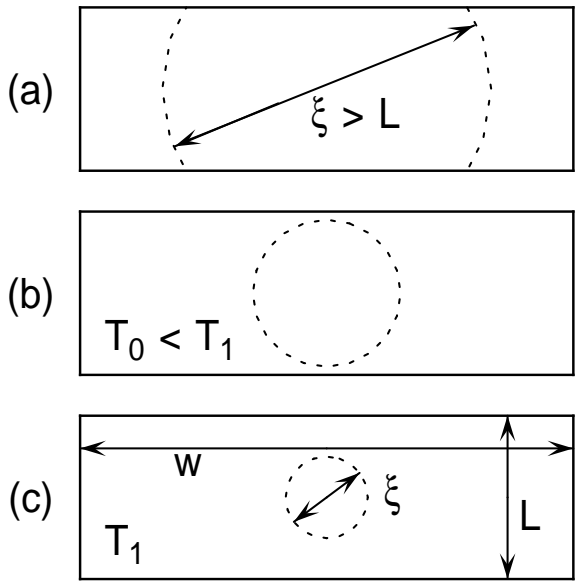,width=3in}}
\vspace{0.2cm}
\caption{
Illustration that the effects of growth of $L_\varphi \propto 1/T^p$
can be cut-off by a finite sample length $L$.
\label{finitesize}
}
\end{figure}

In generic silicon metal-oxide-semiconductor field-effect transistors
(MOSFETs), which comprise strictly 2D electrons (or holes) between
reservoirs of 3D carriers,\cite{AFS} there is a MIT\cite{Krav}
for MOSFETs with sufficiently low disorder and high carrier
mobility\cite{DP_PRL}.  In the MOSFETs, the transition occurs among
normal (non-superconducting) carriers at zero magnetic field.
The transition occurs at relatively low electron densities $n_s$ of 
the order of $1\times 10^{15}/m^{2}$, where the Coulomb energy ($U \sim
\sqrt{n_s}$) is larger than $E_F$.  

Clear signatures of the 2D MIT have been observed at $B=0$ in generic
2DES as a reversal of the sign of the temperature coefficient of
conductance as the carrier density crosses through some critical value
$n_c$: $dG(T)/dT$ changes from positive to negative.  The same
signature is found in a variety of materials including electrons in
Si-MOSFETs,\cite{Krav,DP_PRL} holes in GaAs/AlGaAs
heterostructures\cite{pGaAs,pGaAs_Cam}, holes in SiGe/Si
heterostructures\cite{pSiGe}, and electrons in Si/SiGe\cite{comment}.

For all versions of the 2D MIT, a scaling of the resistivity data
against a control parameter can be accomplished in a form that
directly follows from standard scaling considerations. 
In earlier work, this scaling form has been 
employed to explain the
superconductor-insulator transition in granular metal
films\cite{SIT_Th}.  Written in the form appropriate to describe the
MOSFET experiment (where the control parameter is $n_s$), the scaling 
equation is\cite{newgang}
\begin{equation}
\label{eq1}
\rho (T,n_{s})=f(|\delta_{n}|/T^{1/z\nu})=\rho (T/T_{0}),
\end{equation}
where $\delta_{n} ~ = ~ (n_s - n_c)/n_c$ is a distance from the
transition and $z$ and $\nu$ are exponents associated with the scaling
of a microscopic correlation length $\xi$, which measures the regions
over which the quantum interactions are effective -- {\it i. e.} it is
related to the phase coherence length
$L_\varphi$.  In the vicinity of the MIT, the correlation
length is assumed to scale to infinity as $\xi ~ = ~ 1/|\delta_{n}|^{\nu}$.

Note that standard scaling arguments are typically formulated in terms
of the scale dependence of the conductance at $T=0$.  
It is clear that the $T=0$ limit of the theoretical models is
a poor approximation to any experiment, since it presumes that the
coherence length for the carrier excitations is infinite. As first
emphasized by Thouless \cite{thouless}, this assumption is false at
any finite temperature; more recent arguments \cite{VA_Brownian}
question its validity even at $T=0$.  The predicted $\beta$ can be
related to more useful observables, by noting that the effective
sample size for quantum interactions and interferences is not the
patterned sample size but the length-scale $L_\varphi$ on which
carriers retain phase coherence or phase memory.  For most systems,
this characteristic thermal length scales as a power of the
temperature, viz.  $L_\varphi \propto 1/T^{1/z}$, where $z$ is the
so-called dynamical exponent ($z=1/p$ in  ``weak localization''
parlance).  
In this way, the scaling function in the length ``domain'' can be
converted to $\beta_T = - d(\ln g)/d(\ln T)$ 
in the $T$ ``domain''\cite{newgang}.
We emphasize that if the sample size $L < L_\varphi$, then temperature
dependence may be ``short-circuited'' by the finite size effects as
illustrated in figure \ref{finitesize}.

The use of the quantum phase transition\cite{QPT} models to explain the 2D
MIT carries a certain baggage with it.  Such  models are generally 
accepted for a superfluid (boson) ground state and transition only
at $T=0$ to an insulating phase.  For the quantized Hall effect or the
granular superconductor, there is an obvious choice of ground state in
the dissipationless ($\rho = 0$) behavior characteristic of these two
physical systems.  It begs the question, however, of the ground state
in the high-mobility MOSFETs: What is going on at $T=0$ ?  The
appearance of the 2D MIT in experiments has led to something of an
avalanche of theoretical ideas, with ``explanations'' of the metallic
behavior ranging from valley crossings,\cite{valleys} triplet
superconductivity,\cite{triplet} anyon
superconductivity\cite{annoyon},...  There also have been
suggestions that non-interacting models
with spin-orbit scattering can explain the data\cite{Pudalov}.
These efforts nonwithstanding, the nature of the 2D metallic
ground state and the 2D MIT remain some of the most challenging open 
problems of contemporary condensed matter science. 

\section{Experiment}

We have studied the conductance $G$ of two different sets of
high-mobility n-channel Si-MOSFETs.  A set of large area ($\simeq
1$mm$^2$) FETs were fabricated on the (100) surface of silicon wafers
doped with $N_a \approx 8.3\times 10^{20}$m$^{-3}$ acceptors.  Corbino
channels of length $L=0.4$~mm and width $w=8$~mm were formed with 
a poly-silicon gate above a 44~nm oxide layer.
The measured residual oxide charge for these devices was $\approx
3\times 10^{14}/$m$^{2}$.  These samples have been studied only at
relatively high temperatures $1.2 < T < 4.2$~K.  Another set of
samples made in a different process run on very similar starting
material had a residual oxide charge of $< 10^{14}/$m$^{2}$.  This run
comprised samples of various lengths $1~\mu$m$ < L < 256~\mu$m and
widths $11~\mu$m$ < w < 500~\mu$m.  For the shorter samples $w \gg L$,
so that even though $L$ is small enough to compete with important
microscopic process length scales in the material there is some hope
of inferring averaged properties as a result of having many
``mesoscopic'' elements in parallel.~\cite{FkLS} This later batch of
samples have been studied at much lower temperatures $0.01 < T <
4.2$~K.  All sets of FETs had rather high peak mobilities $\simeq
1$~m$^2/$V$\cdot$sec, or they could achieve this regime with substrate
bias.\cite{DP_PRL}

All measurements of $G$ were conducted in an electrically shielded
enclosure with standard lockin techniques.  A source-drain voltage
$V_{sd}$ was applied and the resulting current $I_{sd}$ was recorded
as a function of the temperature $T$, the magnetic field $B$, or the
gate voltage $V_g$, which controls the carrier density $n_s ~ = ~
C_{ox} (V_g - V_{th})$.  $V_{th}$ is the threshold for populating
the channel, which can be inferred from higher temperature
transconductance or from Shubnikov-de Haas measurements at low
temperatures, and $C_{ox}$ is the specific capacity of the gate oxide.

\begin{figure}
\centerline{\psfig{figure=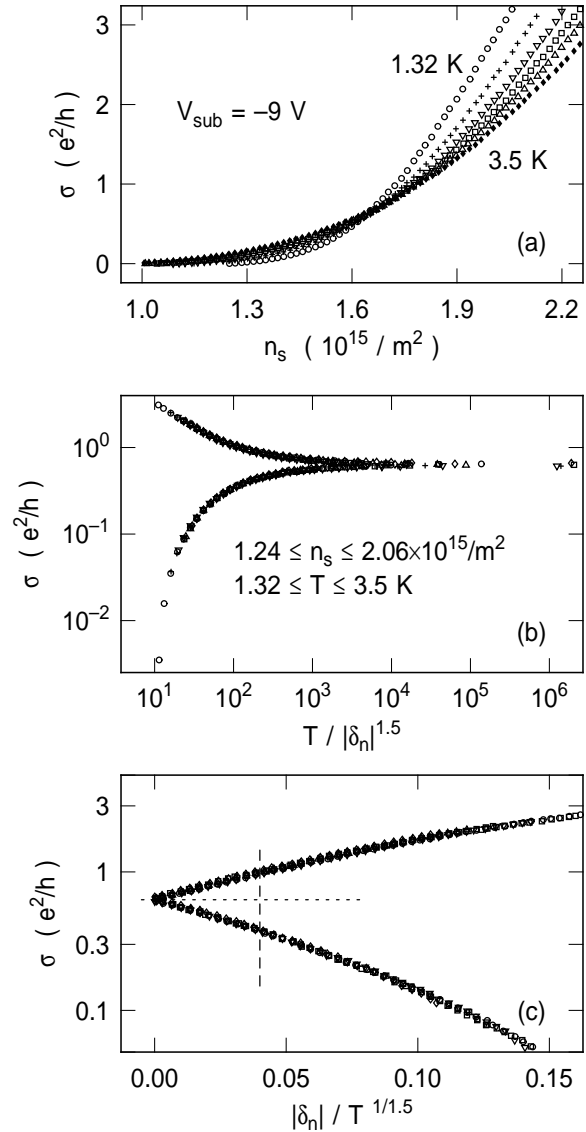,width=3in}}
\vspace{0.2cm}
\caption{
(a) $\sigma(T,n_s)$ of a large MOSFET for $T~=~$ 1.32 ($\circ$), 1.84
($+$), 2.30 ($\bigtriangledown$), 2.70 ($\Box$), 3.00
($\bigtriangleup$) and 3.50 K ($\Diamond$).  
(b) The same data re-scaled according to Eq.~(\ref{eq1}) to form a 
single function with two branches (bottom branch for the insulating 
phase and top branch for the metallic behavior) 
(c) The data re-scaled again in accord with Eq.~(\ref{eq2}) so that 
the symmetry for insulating and metallic
conductance is evident up to $T = T_0$, whch is marked by the dashed
line.  The dotted line marks the conductivity $\sigma_c = 0.8 e^2/h$
at the MIT.  
(These data were recorded at finite magnetic field
$B=0.5$~T, but this has no bearing on the shapes of any of the curves
for this sample.)
\label{dhva5cond}
}
\end{figure}

\section{Results for $B=0$}

Our experiments demonstrate first that the disorder in the devices
must be low relative to other important energy scales.  
For a moderately disordered sample, we find that the
slope of $G(T)$ is consistently positive: all values of
$n_s$ scale as insulators and the conductivity vanishes 
as $T \rightarrow 0$.
For a substantial back-gate bias ({\it e.g.} $V_{sub} = - 9$~V
for the sample shown in Fig.~\ref{dhva5cond}), however, the mobility 
increases enough to permit a MIT at a value of $n_s = n_c \simeq 
1.65\times 10^{15}/m^{2}$.  The MIT appears as a change of sign of 
the slope of $\sigma (T)=(L/w)G(T)$ at $n_c$ (see 
Fig.~\ref{dhva5cond}(a)).  
We infer that the peak mobility has increased by
about 20\%, but this clearly is sufficient to change the
behavior of the 2DES dramatically, even though the value of 
$\sigma$ at $n_{s} = n_{c}$ has not changed much. 

More importantly, all of the conductance curves $G(V_g,T) = G(n_s,T)$
can be scaled (see figure \ref{dhva5cond}(b))
against $\delta_{n} = (n_s - n_c)/n_c$ to form a single
two-branched function [Fig.~\ref{dhva5cond}(b)]
as expected from the scaling arguments mentioned above.  We take this 
as a signature of a quantum phase transition in the 2DES.
In figure \ref{dhva5cond}(c) we plot
\begin{equation}
\label{eq2}
\sigma (\delta_n,T)=\sigma_{c}\exp\left(A \delta_{n}/T^{1/z\nu 
}\right),
\end{equation}
for all data from figure \ref{dhva5cond}(b), and find agreement with the
theoretical prediction~\cite{newgang} for the temperature dependence in
the quantum critical region $T>T_0$, where the crossover temperature 
$T_0\propto |\delta_{n}|^{z\nu}$ is shown by the dashed line in 
Fig.~\ref{dhva5cond}(c).

Very similar scaling of $\sigma$ has been obtained in other 2DES
samples in a variety of physical
circumstances.\cite{SIT_Exp,Shahar,Diorio,Krav,DP_PRL} A related scaling
of $\sigma(E,T,n_s)$, where $E$ is the electric field applied between
the source and drain, has been observed in some
2DES.\cite{Efieldscaling} This collection of experimental results is
overwhelming evidence that the MIT is a quantum phase transition, and
exhibits similarities
with transitions in granular superconducting thin films,
oxide superconductors, and plateaux transitions in the quantized Hall
effect.

\begin{figure}
\centerline{\psfig{figure=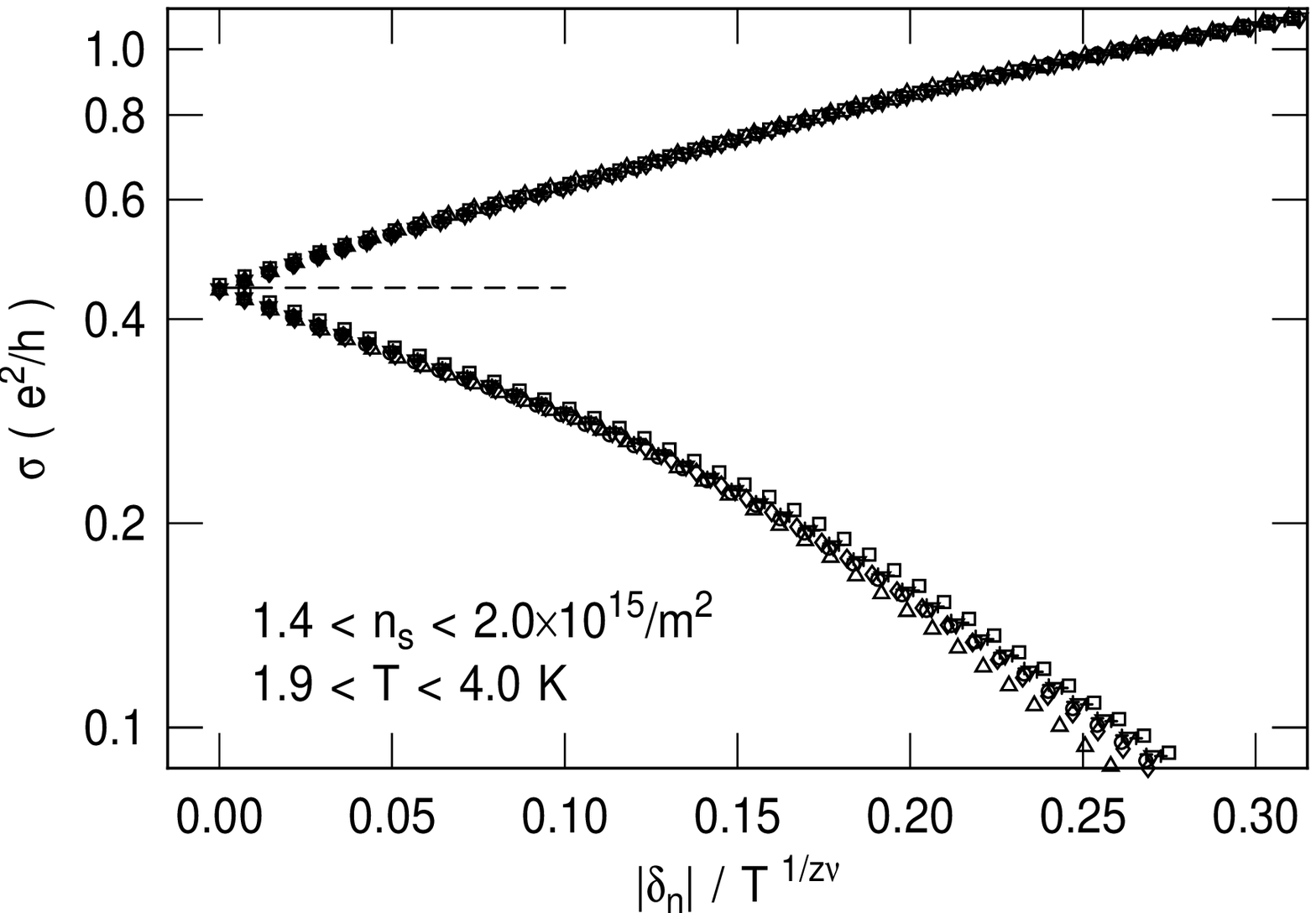,width=3in}}
\vspace{0.2cm}
\caption{
Scaling of $\sigma(n_s,T)$ for a 1.25 $\mu$m long 11.5 $\mu$m wide
MOSFET for $T =$ 1.97 ($\bigtriangleup$), 2.48 ($\diamond)$, 
2.68 ($\Box$), 3.06 ($\times$), 3.53 ($\bigtriangledown$) 
and 3.93~K ($\circ$, $+$).  
The dashed line marks $\sigma_c = 0.45 e^2/h$.
\label{KPLscaling_plot}
}
\end{figure}

In shorter MOSFETs we find a similar scaling of $\sigma$ for moderate
temperatures $1 < T < 4.2$~K.  Our results are illustrated in figure
\ref{KPLscaling_plot} which contains measurements from 
a $L = 1.25~\mu$m, $w = 11.5~\mu$m MOSFET.
The reversal of sign of the slope $dG/dT$ occurs at essentially the
same value as for the larger MOSFETs, and the conductivity $\sigma_c$
at the MIT is $\sim e^2/2h$ in agreement with many other experiments
on MOSFETs.  A similar value of $\sigma_c$ obtains in completely
different physical circumstances such as the granular films or the
quantized Hall systems, but there the amplitude of $\sigma$ is
governed by physics (charging energy and phase fluctuations or the
dominance of a particular momentum mode) that are not directly
germane to the MIT.

\begin{figure}
\centerline{\psfig{figure=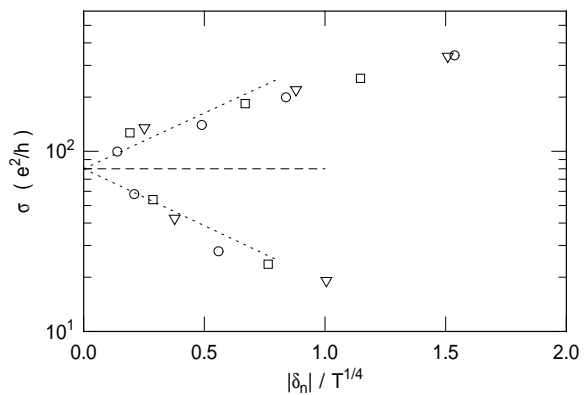,width=3in}}
\vspace{0.2cm}
\caption{
Scaled $\sigma(n_s,T)$ for $T=4.2, 1.2$ and 
0.4~K; $V_{sub}=10$~V; $n_{c}\approx 1.8\times 10^{15}/$m$^{2}$.  
The dashed line marks $\sigma_c = 80 e^2/h$, and the dotted lines guide the
eye.
\label{commentfig}
}
\end{figure}

Yet a third set of samples with electrons confined at the interface
between Si and SiGe has exhibited very similar response. \cite{comment}  
This result (see figure \ref{commentfig})
is especially intriguing, because the mobility is very high compared
with the MOSFETs experiments~\cite{SiGe} -- in the same range as for 
electrons in GaAs heterostructures, which have not exhibited the 
MIT to date.  
Again the slope $dG/dT$ changes sign at a concentration $n_c$ =
1.8$\times 10^{15}/m^2$, but owing to the higher mobility, the
value of $\sigma_c \simeq 80e^2/h$ is {\it very} large compared to the
MOSFET experiments.  This result proves the existence of the MIT in
yet another materials system.  More interestingly, it lays to rest once
and forever the suggestion (quoted widely and believed even more
widely) that the value $\sigma_c$ is a ``universal'' number.  
Indeed, subsequent studies of the MIT in a 2D hole systems in 
SiGe/Si~\cite{pSiGe}
and GaAs/AlGaAs heterostructures~\cite{pGaAs_Cam} have found
$\sigma_{c}\sim 2e^2/h$, almost a factor of seven larger than in some
Si MOSFETs.

A much more plausible expectation is that 
$r_s=U/E_F$ will be a universal number.  This, in fact,
is not born out across all samples: 
$r_{s}\sim 12-19$ at the transition for electrons in Si MOSFETs and 
Si/SiGe quantum wells,
$r_{s}\sim 13$ for holes in SiGe/Si heterostructures and 
$r_{s}\sim 11-23$ for holes in GaAs/AlAs heterostructures.  These numbers,
however, are much closer to each other (within a factor of two) 
than the values of $\sigma_c$, which are
distributed across more than two orders of magnitude.

In general, one expects the scaling exponents, such as z and $\nu$, to
reflect only the symmetry of the problem in question, and thus to be
universal numbers.
In the experiments, however, a large range of exponents are found.  In the
Si-MOSFETs at $B=0$, $z\nu=1.6\pm 0.2$ has appeared in two different
sets of samples.\cite{Krav,DP_PRL}  
The same exponent has been reported for holes in
SiGe/Si quantum wells.\cite{pSiGe}  For holes in GaAs/AlAs heterostructures 
\cite{pGaAs_Cam} or
electrons in Si/SiGe heterostructures \cite{comment}, however, $z\nu > 4$.  
It is
worth noting that these two experiments were on samples with very low
disorder ($k_F l \gg 1$) by comparison with the MOSFET and p-SiGe
samples where the relative effect of disorder is much stronger ($k_F l
\simeq 1$).  
In the very short MOSFETs\cite{Li,DP_RC}, $z\nu$ is even larger
-- $z\nu = 16 \pm 4$ for $L \approx 1 \mu$m.  Furthermore, the range
of ``scalable'' temperature dependence is smaller than in the larger
samples.  
In fact for $|\delta_n| > 0.15$ on the 
insulating side, the lowest temperature 
curve ($\bigtriangleup$) is
already failing to scale with the higher temperature data.
In the sample of figure \ref{dhva5cond}, the data obeyed
Eq.~(\ref{eq1}) well for $1.2 < T < 3.6$~K ($T=1.2$~K was the lowest
temperature studied in that experiment), and for other large
MOSFETs\cite{Krav} the range of temperature extended down to 0.05~K.
We attribute this saturation of the $T$ dependence to the cut-off of
$L_\varphi$ and hence $\xi$ by the sample length $L$.

\section{Magnetoconductance}

An applied magnetic field $B$ alters the behavior of the carriers in
the 2DES in different ways depending on the angle between the field
and the plane of the 2DES.  The perpendicular component of the
magnetic field tends to bend the path of the carrier (the Lorentz
force tends to make the carriers form circular orbits), as well as
inducing a splitting of the energies of different spin orientations
and consequently to polarize the spins of the carriers out of the
plane of the 2DES.  A parallel component of the magnetic field
probably has little effect on the orbital motion of the carriers, but
it still splits spin energies and it polarizes the spins into the
plane.  Studies of $G(B)$ have been employed fruitfully for decades as
probes of the transport mechanisms of disordered
systems.\cite{LR} There are detailed calculations of the
functional form of $G(B)$ for various conditions among non-interacting
carriers\cite{LR}.  The contributions from Zeeman interactions
(triplet terms in Hartree approximation) have been calculated and for
$\mu g B$ \lap $k_BT$ ($\mu g$ is the carrier magnetic moment and
$k_B$ is Boltzmann's constant) lead to a negative parabolic
magnetoconductance in a perpendicular magnetic field.  In contrast,
quantum interference of carriers (orbital effects) can lead to
positive slopes of $G(B)$ near $B=0$.

The MIT in MOSFETs is quenched by a large parallel magnetic field ($B$
\gap 1~T)\cite{Kravbpar,B_parallel}, which in turn implies that polarizing 
the electron spins {\it in the plane of the carrier motion} has a dramatic
effect on the correlations of the carriers.  The conductance of
nominally metallic 2DES ({\it i.e.} $n_s > n_c$) decreases by orders of
magnitude and the slope of $G(T)$ reverses from negative to positive,
indicative of insulating samples.  The application of $B$ perpendicular 
to the plane of the 2DES 
also quenches the metallic phase~\cite{DP_unpubl,Krav_tilted}.
Moreover, careful measurements of magnetoconductance in the quantum
critical region in the presence of a perpendicular $B$ 
provide quite a bit of insight into the relative importance of spin
interactions and orbital motion at the MIT.\cite{DP_PRL}

\begin{figure}
\centerline{\psfig{figure=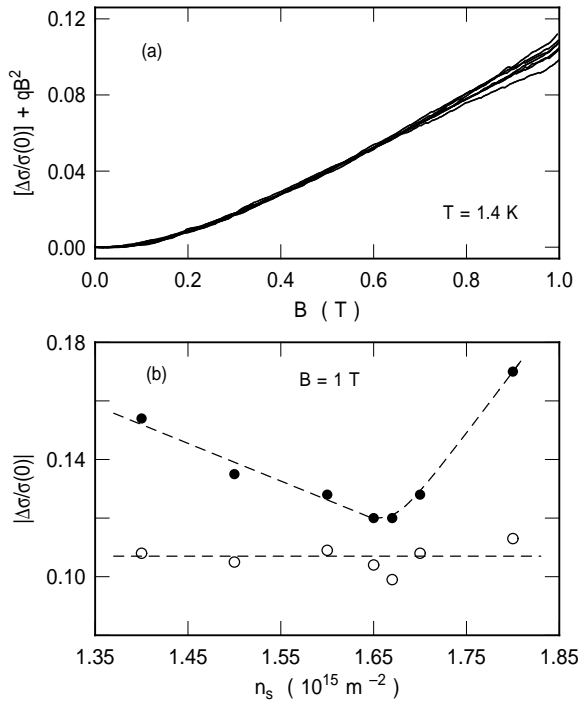,width=3in}}
\vspace{0.2cm}
\caption{
(a) The positive contribution $p f(B)$ 
to $\Delta \sigma/\sigma(B=0)$ 
for all $1.40 \leq n_s \leq 1.80~\times 10^{15}/$m$^2$.  
(b) Amplitudes $p$ ($\circ$) and $q$ ($\bullet$)
of $\Delta \sigma/\sigma(B=0)$ as a function $n_s$ for $B=1$~T.  The
lines guide the eye.
\label{dhva1mc2}
}
\end{figure}

Figure \ref{dhva1mc2} contains representative plots of 
magnetoconductance (MC) $\Delta
\sigma/\sigma(B=0)$, where $\Delta\sigma = \sigma(B)-\sigma(B=0)$.  
Clearly, there is a positive magnetoconductance and a negative
magnetoconductance contribution to each curve, and the positive
contribution is more accentuated near $n_s = n_c = 1.65 \times
10^{15}/m^{2}$.  As we will see below, this is because the negative
contribution has dropped to a minimum value at this point.  
\begin{figure}
\centerline{\psfig{figure=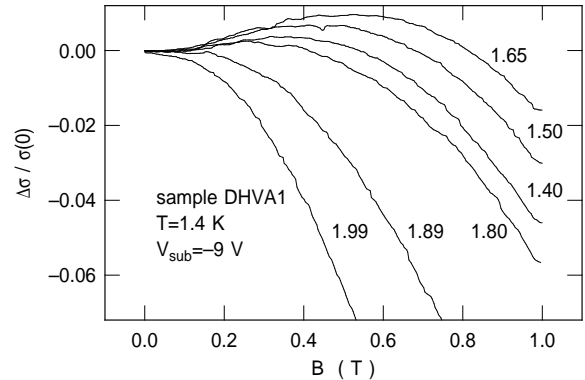,width=3in}}
\vspace{0.2cm}
\caption{ 
Magnetoconductance for a large MOSFET with $V_{sub}=-9$~V and
$n_{c}=1.65\times 10^{15}/$m$^{2}$.  
The different curves arise from the values of $n_s$ 
written on the plot (units of $10^{15}/$m$^{2}$).  
\label{dhva1mc}
}
\end{figure}

Each curve in figure \ref{dhva1mc} can be written
\begin{equation}
\label{eq3}
\Delta\sigma /\sigma (0) ~ = ~ p f(B) - q B^{2}, 
\end{equation}
where $p$, $q > 0$.  The assumption of parabolic form for the negative
term is justified because (1) it fits the data and (2) it is the form
expected for the electron-electron interaction contribution for $B$
\lap $k_BT/\mu g \simeq 0.8$~T~\cite{gabriele}.  We have decomposed 
the data following formula 
(\ref{eq3})and find that the positive component $p f(B)$ of the
magnetoconductance does not depend on $n_s$ within the scatter of
data.  These
contributions are plotted in figure \ref{dhva1mc2}(a) and fall atop
one another for $B<0.8$~T.  The coefficient $p$ is therefore a
constant.  The coefficient $q$, {\it i.e.} the electron-electron interaction
contribution to MC, exhibits a minimum near $n_s=n_c$ as is
obvious in figure \ref{dhva1mc2}(b).  
We have found exactly the same behavior of MC in short 
MOSFETs~\cite{Li,KPLi_unpubl} at temperatures down to 0.04~K,
in spite of the presence of sizable conductance fluctuations.

Both positive and negative MC have been observed in other MOSFETs as
well~\cite{B_parallel}.
Since orbital effects (positive magnetoconductance) are absent in an
applied parallel field, our results show clearly that MC of a 2DES
{\it does} depend on the angle between the magnetic field and the plane
of a 2DES, at least in the quantum critical region.  Away from the
critical region, the orbital effects at low fields become small 
compared to spin effects [see Fig.~\ref{dhva1mc2}(b)], and only negative
MC is observed.  Therefore, a recent claim~\cite{Krav_tilted} about
the absence of any angular dependence of MC is valid only far from
the critical regime, in the metallic phase where the 
experiment~\cite{Krav_tilted} was carried out.

\section{Scaling of the conductance}

\begin{figure}
\centerline{\psfig{figure=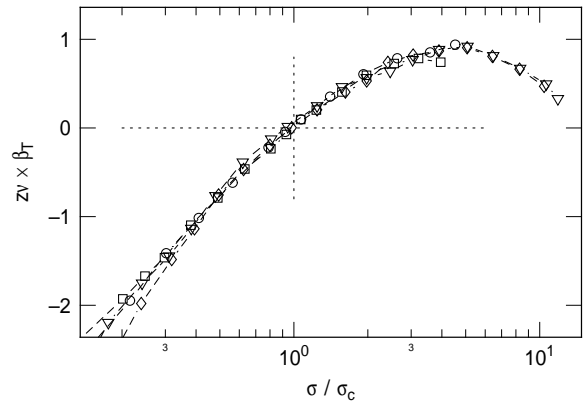,width=3in}}
\vspace{0.2cm}
\caption{
Experimental values of $\beta_T$ scaled by $1/z\nu$ for four samples:
two large area ($\bigcirc$ and $\Box$) MOSFETs and two short
($\bigtriangledown$ and $\Diamond$) samples (for high enough $T$).
\label{betafunctn}
}
\end{figure}

We can extract $\beta_T \propto \beta$ from our data by using the
assumption of power-law dependence of the correlation scale (the phase
coherence length) on $T$.\cite{LR,newgang}
One such plot for four of our samples is given in figure \ref{betafunctn}.
Anticipating the dependence of $L_\varphi$ suggested in the introduction, 
we have 
scaled the bare $\beta_T$ multiplying it by $z\nu$.  The two large MOSFETs 
have $z\nu = 1.6$, and the two short MOSFETs yielded $z\nu = 16$.
In spite of the disparity,
all four curves produce the same scaled scaling function indicating
that the underlying $\beta$ is the same for all samples.  A
noteworthy feature is that $\beta$ is essentially linear as it
crosses through $\beta = 0$ as expected from recent scaling arguments
\cite{newgang}.  For non-interacting particles, $\beta < 0$ as illustrated 
by the dotted lines in figure \ref{betaplot}.  Our experiment (which
corroborates other measurements\cite{Krav_beta}) proves that
$\beta$ extends into the metallic region and thus provides clear
evidence as to the inadequacy of noninteracting models of charge carriers.

An even more striking feature is evident from the
present data: $\beta$ decreases towards zero
at large $g$, as proposed in Ref. \cite{newgang}. Although this is by
no means a necessary condition for such an exotic metallic state, it
seems as if the system tries to restore Ohm's law in the large
conductance (weak disorder) limit. 

\section{Summary}

Our experiments corroborate other experiments on 2DES and prove the
existence of a metal-insulator transition in two-dimensions for
sufficiently low disorder.  The conductances can be reduced to a
single (two-branched) scaling function that is consistent with recent
proposals for Coulomb driven quantum phase transitions.  Measurements
for different sample lengths and different levels of disorder
demonstrate certain universal features of the transition.  The
transition occurs near a certain value of $n_c$ for a given density of
states (carrier effective mass).  The conductance at the transition is
{\it not} a universal constant.
Magnetoconductance
experiments suggest that spin polarization and Coulomb interactions of
the carriers have a dramatic effect on the transition.  The product of
conductance scaling function $\beta_T$ and the correlation scale
exponents $z\nu$ shows that the underlying scaling function $\beta$
is the same for all MOSFETs.  The inferred $\beta$ violates the
predictions from non-interacting models and supports recent
predictions based on interacting carriers. 

\section{Acknowledgment}
The authors are grateful to V. Dobrosavljevi\'{c} for
useful discussions.  This work was supported by NSF Grant
No. DMR-9510355 and ARO Grant EL-33036.


\end{document}